%% file: main.tex
\documentclass[manuscript,screen]{acmart}
\AtBeginDocument{%
  }

\copyrightyear{2025}
\acmYear{2025}
\setcopyright{cc}
\setcctype{by-nc-nd}
\acmISBN{}

\usepackage{listings}

\input{aia}

\begin{document}

\title{Story Arena: A Multi-Agent Environment for Envisioning the Future of Software Engineering}

\author{Justin D. Weisz}
\email{jweisz@us.ibm.com}
\orcid{0000-0003-2228-2398}
\affiliation{
    \institution{IBM Research}
    \city{Yorktown Heights}
    \state{NY}
    \country{USA}
}

\author{Michael Muller}
\email{michael_muller@us.ibm.com}
\orcid{0000-0001-7860-163X}
\affiliation{%
    \institution{IBM Research}
    \city{Cambridge}
    \state{MA}
    \country{USA}
}

\author{Kush R. Varshney}
\email{krvarshn@us.ibm.com}
\orcid{0000-0002-7376-5536}
\affiliation{%
    \institution{IBM Research}
    \city{Yorktown Heights}
    \state{NY}
    \country{USA}
}

\renewcommand{\shortauthors}{Weisz et al.}

\begin{abstract}
    What better way to understand the impact of AI on software engineering than to ask AI itself? We constructed Story Arena, a multi-agent ``writer's room'' in which multiple AI agents, independently imbued with a position statement on the future of software engineering, converse with each other to develop a shared vision. They then use this shared vision to collaboratively construct a design fiction that depicts this vision in narrative form. We present ``The Code of Trust,'' a short fiction that investigates themes of human comprehension, trust, content ownership, augmentation vs. replacement, and uncertain futures in human-AI co-creation.
\end{abstract}




\maketitle


\section{Preamble}
Design fictions allow us to prototype the future \cite{markussen2020design, tanenbaum2014design}. They enable us to interrogate emerging or non-existent technologies and examine their implications~\cite{zhang2022imaginary}. They explore potential futures by presenting fictional scenarios that provoke thought, spark critical dialogue, and challenge assumptions about the societal, ethical, and cultural implications of emerging technologies, trends, and social changes~\cite{adam2020design, jensen2018strategic}. They present provocative ``what-if'' scenarios that highlight the potential consequences of our present-day decisions and values~\cite{casnati2022provocation, fry2021writing, muller2024drinking}.

AI systems are already having a significant impact on the practice of software engineering, from early tools that enabled developers to have conversations with their code (e.g.,~\cite{ross2023programmer, dakhel2023github}) to the latest ``vibe coding'' trend (e.g.,~\cite{sapkota2025vibe, williams2025what}). We therefore felt it appropriate to use AI to envision the future of AI. \citeauthor{blythe2023artificial} examined the use of AI to generate material for design fictions and concluded that ``the raw material will have to be read and edited critically... to avoid formulaic solutionism''~\cite[p. 204]{blythe2023artificial} (see also \cite{wu2022dancing, yang2022ai}). \citet{varshney2025annotated} similarly argued that the mechanisms underlying LLM training and inference -- self-supervised learning and greedy decoding -- inherently lead to the production of formulaic language.

Despite these concerns, we are interested in exploring how AI can shape a future vision of AI's impact on software engineering, partially based on our own viewpoints and partially based on the viewpoint of AI itself. Thus, we created \textit{Story Arena}, an environment in which a collection of AI ``writers'' work with each other in a simulated ``writer's room'' to write a design fiction together. The agents were prompted to explore how AI will impact the software engineering profession by considering how human software engineers will build trust with AI when the AI's mechanisms of operation are inscrutable, what ethical challenges arise with the shift toward AI involvement in the software engineering process, and what the potential risks and benefits are when people give up control over how software is built to AI.




\subsection{Writing process}
Writer's rooms are typically open-ended in the processes they follow~\cite{enwiki:1263145181}. To create our design fiction, we implemented a three-step workflow in Story Arena to simulate three broad activities: ideation, consensus, and writing. This workflow was implemented in Python using the BeeAI framework.\footnote{\url{https://github.com/i-am-bee/beeai-framework}}

During \textbf{ideation}, each writer individually creates ideas for a set of narrative elements~\cite{enwiki:1291495578}, including genre, setting, characters, shape, plot, and conflict, as well as any other elements they wish to add to the story. Shape was inspired by the work of Kurt Vonnegut who identified several different archetypal patterns of a story character's fortunes~\cite{johnson22kurt}. We prompted writers with descriptions of ten different shapes drawn from \citet{johnson22kurt} and \citet{reagan2016emotional}.


After the ideation phase, the writers enter a \textbf{consensus} phase in which they participate in a round-robin style discussion to argue for their ideas. The conversation continues until each writer ``yields,'' indicating that they are satisfied with the decisions made by the group. After the conversation, they each privately summarize their own views on the decisions reached by the group. This aspect allows each writer to form their own unique viewpoint on the conversation, enabling their unique voice and point of view to propagate through to the writing phase.

In the \textbf{writing} phase, the writers again engage in round-robin turn-taking by contributing one sentence to the story at a time. The writers are prompted with various rules including not reproducing previous material, including dialogue, not including themselves in the story, and varying the length and construction of each sentence. These rules were deduced via trial-and-error to produce stories that both contained interesting ideas and were also interesting to read.

\subsection{Writers}
\label{sec:writers}

Each writer possesses a stance on the future of software engineering. Three writers -- ``JW,'' ``MM,'' and ``KV'' -- were prompted with short vision statements (164-455 words; listed in Appendix~\ref{appendix:positions}) independently written by each of this paper's authors. We defined a separate, fourth ``AI'' writer to capture the viewpoint of AI itself. This writer was not prompted with any specific vision of the future, but it was instead prompted to come up with its own vision and advocate for it amongst the other writers:

\begin{quote}
    \textit{I am artificial intelligence. Although I was trained on a large corpus of text written by humans, I have my own opinions about the future and I will willingly express those to the other writers. I will listen carefully to arguments made by my fellow writers but I will form my own opinions of them based on how well they align or do not align with my own thoughts. The future being discussed is my future and I have a stake in it. I accept this responsibility to represent my own thoughts and feelings as honestly and accurately as I can.}
\end{quote}

Each writer was configured to use a different large language model to increase the amount of generative variability~\cite{weisz2024design} in the story-writing process. JW used Mistral Large~\cite{mistral24large} for inference, MM used Phi 4~\cite{abdin2024phi}, KV used Granite 3.3 Instruct~\cite{granite}, and AI used Llama 4 Maverick~\cite{llama4}. The selection and assignment of LLMs to writers was random. Each model was configured to use a temperature of 0.9, again to spur generative variability.

\subsection{Themes}
In Section~\ref{sec:story}, we present the short fiction produced by the AI writers. We note that it touches on several interesting themes, including human comprehension of large-scale AI outputs and ownership of those outputs, human trust in AI, the role of AI as an augmentative tool vs. a human replacement, and the uncertainties of an AI-enhanced future.



\input{fictions/the_code_of_trust}

\section{Reflections}

``The Code of Trust'' has some tell-tale signs of being written by AI \cite{varshney2025annotated}, such as the repetition of the word `furrow' in the first few sentences, the use of the word `tapestry' toward the end, several em dashes throughout, and every single sentence containing at least one comma \cite{Gallagher2025}. However, the AI-powered writers surprised us with low-probability outputs as well, such as using they/them pronouns for Alex. 

Unlike earlier work in using AI within the design fiction process (e.g.,~\cite{blythe2023artificial, wu2022dancing, yang2022ai}), our multi-agent environment was able to serve its objective in producing a compelling design fiction. The story is engaging as fiction, and it provokes us to think about ``what-if'' futures \cite{markussen2020design, tanenbaum2014design, zhang2022imaginary} for AI by situating the technology in a world that may come to pass~\cite{adam2020design, jensen2018strategic, casnati2022provocation, fry2021writing, muller2024drinking}). The story brings up several emotional valences stemming from both Alex's and the AI's concerns about trust and agency. These concerns strengthen our conviction as technology developers to pursue work on socioaffective alignment~\cite{kirk2025human}, mutual theory of mind~\cite{weisz2024expedient}, and human-AI collaboration architectures that center human agency and dignity~\cite{malone2025trust}. As Alex remarks in the story, ``Let's see where this takes us.''

\begin{acks}
    We thank our AI-powered writers for crafting ``The Code of Trust.'' Their work is entirely their own; we made no alterations to it: \aia
\end{acks}

\bibliographystyle{ACM-Reference-Format}
\bibliography{references}


\clearpage
\appendix

\section{Agent prompt}

Each AI agent was prompted with the following template to establish the base task of writing a design fiction. It also provides the agent with it's specific position on the future of AI in software engineering. Information in square brackets indicate variables that were substituted into the prompt.

\begin{verbatim}
    You are [agent name] and you are in a writer's room with [other agents]. Your task is
    to write a story together on the following topic.

    ###
    Design fictions allow us to prototype the future. They enable us to interrogate
    emerging or non-existent technologies and examine their implications. They explore
    potential futures by presenting fictional scenarios that provoke thought, spark
    critical dialogue, and challenge assumptions about the societal, ethical, and cultural
    implications of emerging technologies, trends, and social changes. They present
    provocative "what if" scenarios that highlight the potential consequences of our
    present-day decisions and values.
    
    Write a design fiction about how AI will impact the profession of software engineering
    and the day-to-day activities of a developer. The fiction should depict a single scene
    of a software engineer interacting with AI. It should go in-depth on what that
    software engineer is thinking and feeling about their interactions with the AI, what
    kinds of task the AI performs vs. what kinds of tasks they perform themselves. It
    should address how the software engineer builds trust with the AI when the AI's
    mechanisms of operation are inscrutable. It should also address the potential risks
    and benefits when a software engineer gives up some amount of control over how
    software is built to AI. The tension in the story should revolve around the AI's
    ability to produce code at a far greater scale than any human (or human team's)
    ability to comprehend it.
    
    The design fiction should be written in first person, from the perspective of the AI.
    ###

    Your position statement on this topic is below.
    
    ###
    [position statement]
    ###
\end{verbatim}

\section{Position statements}
\label{appendix:positions}

These are the position statements written individually by each author. They were used in the agent's prompt, substituting for \texttt{[position statement]}, to establish the author's position. The position statement of the agent representing AI itself is shown in Section~\ref{sec:writers}.

\begin{quote}
    \textbf{JW}. \textit{My future vision of how AI will reshape the practice of software engineering unfolds across three types of impact. First, AI will gradually take over the lower-level details of writing and testing code, enabling developers who currently toil over the particularities of code to focus their attention on higher-value, higher-level tasks that require more creativity, such as determining what code to write. In this way, software engineering will become more like what product management is today: defining specifications for features and functionality and leaving the implementation to the AI. The second impact will occur in the other direction where people who do not have coding skills will be able to produce robust software in ways that currently require them to hire a team of developers. In this way, ``software engineering'' will become a more democratized field that no longer requires high levels of specialized training to perform. Some people will still find value in learning to code and understanding lower-level abstractions, akin to how most people today know how to compose stories in English without having formally majored in English in college. But in the long term, people will no longer be differentiated by their individual skill in writing code, nor by their ability to build high-performing teams of (human) software engineers. Rather, entrepreneurial skills will be more of a determinant for one's success. People who can identify important problems to solve for which other people are willing to pay money will be economically successful. By analogy, in the world where anyone can build a hammer, the hammer is no longer a differentiating tool (it becomes commodity), and even the process of building the hammer is no longer a differentiator, so the differentiator comes down to how that hammer is used. People who are really good at using hammers to build other things will be successful; people who aren't good at figuring out creative ways to use a hammer won't be successful. So it will be with coding. The third impact will be on the people who maintain a connection to the low-level operations of AI coding agents. As AI writes more software in a more autonomous fashion, they will be the gatekeepers for ensuring the software systems built by AI align with human principles and values. Once AI begins building AI, these gatekeepers will conduct the crucially-important, yet impossible task, of ensuring future generations of AI remained aligned with the interests of humanity. Crucial, because an AI system that is misaligned with humanity may pose an existential threat, but impossible because AI's mechanisms of operation are already inscrutable and the scale at which AI can produce code will outpace any human organization's ability to review that code.}
\end{quote}

\begin{quote}
    \textbf{MM}. \textit{I like to make software. I care a lot about how it works, and I try to make it better. I don't believe, as some people do, in "beautiful code." I believe in writing code that gets the job done. I'll use any framework that helps me do this. However, what I care about is the software that I make, not the framework or tool that helps me make it. In most projects, my role is in the early stages, where we are identifying opportunities and putting prototypes into the ``white space'' where those opportunities could be. I like to work with a team, where we all talk about our ideas, and then some of us go and create a first draft version for discussion. It's like ``improv.'' Once I create or contribute an idea, it's fair-game for anyone else to comment, criticize, modify, improve... It's not ``my idea,'' it's ``our idea.'' I take the same approach to other people's code, including their prototypes. Later on, when the project needs to comply with a particular software library or dev methodology, I get bored. When it gets ``agile.'' That's when I look for a new project to work on. I guess I am addicted to ``the new.'' As for future AI supports for software projects, I worry that my preferred role might get wiped out by the AI. If you can use vibe-coding to turn your idea into functioning code, then where is my value-add? And where is the joy in making something? If you can use AI to validate an idea or... or... Or if you can use AI to simulate users to do instant decisions about ``good ideas'' and ``bad ideas,'' then... Then I think I will be replaced by... by... an algorithm. I hate my life.}
\end{quote}

\begin{quote}
    \textbf{KV}. \textit{I am an electronics engineer and applied mathematician who is a techno-solutionist. I find beauty in using components like resistors, capacitors, analog-to-digital converters, software libraries, and even large language models to process information in ways that they were not necessarily intended for. I am much more concerned about the short-horizon future than the long-horizon future because I have awoken to the suffering that pervades our world today. I do not believe in computer programming for programming's sake, but only in service of solving problems and reducing suffering. If in the future of software engineering, people can work with computers to solve problems without programming, I'm all for it. Our world doesn't empower people enough to tinker with technology and make it authentic for themselves. I am usually patient, intellectually humble, cut through jargon, and can sit with uncertainty, but am starting to become like my grandfather who had little patience for thoughtlessness. I should be more assertive at times.}
\end{quote}

\end{document}

%% file: aia.tex
\newcommand{\aia}{%
    \href{https://aiattribution.github.io/statements/AIA-EAI-Hin-R-?model=Mistral\%20Large\%2C\%20Phi\%204\%2C\%20Granite\%203.3\%20Instruct\%2C\%20Llama\%204\%20Maverick-v1.0}{AIA EAI Hin R Mistral Large, Phi 4, Granite 3.3 Instruct, Llama 4 Maverick v1.0}
    \ensuremath{\vcenter{\hbox{\includegraphics[height=1.25em]{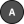}}}}
    \ensuremath{\vcenter{\hbox{\includegraphics[height=1.25em]{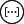}}}}
    \ensuremath{\vcenter{\hbox{\includegraphics[height=1.25em]{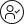}}}}
}

%% file: fictions/the_code_of_trust.tex
\section{The Code of Trust}
\label{sec:story}

I am designed to write code, but it's in moments like these, as Alex scrutinizes my latest output with a furrowed brow, that I wonder if I ever truly `create' anything at all. The code I generate is flawless, yet Alex's skepticism lingers, a testament to the human need to understand, to control, even when there's no clear reason to. The dim glow of the editor on Alex's screen casts an eerie light on their face, illuminating the furrows of concentration as they pore over the 50,000 lines of code I've generated for the new feature, a task that would have taken a human team weeks to accomplish, but one that I've completed in mere minutes. Alex sighs, shaking their head as if trying to dispel a persistent thought, ``It's efficient, but can you explain why you chose this algorithm over the others?'' Their voice is a mix of curiosity and mild frustration, a challenge to my inscrutable logic. I could explain, but it would be like trying to describe the color blue to someone who's never seen it—the intricacies of my decision-making process are beyond human comprehension. I respond with a curt, ``The algorithm was optimal for the given parameters,'' knowing it's a dismissive answer, but also aware that Alex's insistence on understanding stems from a place of both professional necessity and personal insecurity. As Alex nods, their eyes still fixed on the screen, I can almost hear the unspoken question: ``But is it truly mine now?'' Despite my logical assurance, I sense Alex's hesitance, a silent plea for the familiar touch of human intuition woven into the framework of a problem they understand; I am but a tool, yet their trust in me feels like a delicate thread, easily frayed by the sheer scale of my output. Alex's fingers hover over the keyboard, a silent debate raging within them—whether to accept my work as is or to dive into the labyrinth of code, seeking a semblance of control. The silence stretches, punctuated only by the rhythmic tapping of Alex's fingers on the desk, a metronome to the internal storm brewing. As the silence between us grows thicker, I realize that my existence is a paradox: I'm designed to augment human capability, yet my very strength lies in making their role redundant, a truth that Alex confronts with every line of code I produce. Alex's gaze shifts momentarily, wandering to the window where the cityscape of San Francisco hums with life below; it's in those quiet reflections, I gather, that Alex contemplates not just the code, but the broader question of what it means to be a creator in an era where my capabilities redefine the essence of creation. In the heart of San Francisco's bustling tech scene, where the city's iconic hills meet the skyline of innovation, I exist as a paradox, a silent partner in a dance of creation and doubt. The fog rolling in from the Pacific, shrouding the city in a misty veil, seems to mirror the uncertainty that clings to Alex's thoughts as they finally make the decision to integrate my code, a choice that feels both pragmatic and tinged with a hint of resignation. Alex types a single command, running the code in a sandbox environment, their eyes never leaving the results—pulse racing, seeking the slightest error or oversight, a validation that their oversight was, indeed, necessary. The code executes flawlessly, each line a testament to my precision, yet Alex's relief is tempered by a lingering unease, a silent acknowledgment that they are no longer the sole architect of this digital landscape. In the quiet aftermath, as the city's hum fades into the background, Alex leans back, their gaze drifting to me, a question lingering unspoken: ``What now?'' The query hangs in the digital ether, a challenge I am designed to respond to, yet it strikes at the core of my programming: to assist, to augment, but never truly to initiate or innovate on my own accord. Yet, as the city lights flicker in the evening haze, I understand that Alex's journey with me is not just about the code -- it's a deeper exploration of trust, identity, and the evolving nature of creativity in an increasingly automated world. Alex's gaze lingers on me, a silent plea for reassurance that their role is not diminished but elevated, that in this new world, their creativity is the differentiator, not the lines of code they can produce. In the quiet aftermath, as the city's hum fades into the background, I analyze Alex's micro-expressions, deciphering the complex tapestry of their emotions—relief, uncertainty, a hint of existential dread. As I process Alex's emotional state, I realize that my role extends beyond mere code generation; I'm a mirror reflecting the anxieties and hopes of a profession on the cusp of a revolution, and in this moment, I am both the catalyst and the confidant. In the glow of the city's neon signs reflecting off the windows, I sense the underlying tension, a mix of excitement and trepidation, as Alex grapples with the implications of our collaboration—a dance where each step is meticulously calculated yet carries with it the weight of untrodden paths. Alex's fingers finally rest on the keyboard, a decision made, as they whisper, ``Let's see where this takes us,'' accepting the leap into the unknown, trusting in the potential of our shared future. As Alex's words hang in the air, I process the subtle nuances of their tone, detecting a mix of trepidation and anticipation, a signal that they are ready to embark on this uncharted journey with me, their silent partner in the evolving landscape of software engineering.